\documentclass[mnsc,nonblindrev]{informs3-ssrn}
\RequirePackage{tgtermes}
\RequirePackage{newtxtext}
\RequirePackage{newtxmath}
\RequirePackage{bm}
\RequirePackage{endnotes}

\OneAndAHalfSpacedXI 

\usepackage[utf8]{inputenc}

\usepackage{algorithm}
\usepackage{algpseudocode}
\usepackage{amsmath}
\usepackage{amssymb}
\usepackage{subcaption}
\usepackage{array}
\usepackage{todonotes}
\usepackage{longtable}
\usepackage{tikz}
\usepackage{pgfplots}
\usepackage{titlesec}
\usepackage{hyperref}
\usepackage{mathtools}
\usepackage{commath}

\titlespacing*{\section}
{0pt}          
{1.5ex plus 1ex minus .2ex} 
{1.5ex plus .2ex}           

\usepackage{xcolor}

\usepackage{natbib}
 \bibpunct[, ]{(}{)}{,}{a}{}{,}%
 \def\bibfont{\small}%
\let\cite\citep

\EquationsNumberedThrough    

\TheoremsNumberedThrough     
\ECRepeatTheorems  %

\MANUSCRIPTNO{}

\begin{document}

\makeatletter
\renewcommand{\thefootnote}{\fnsymbol{footnote}}
\makeatother
\setcounter{footnote}{1}


\RUNAUTHOR{Zeng and Umrawal}


\RUNTITLE{Mood-Aware Music Recommendation: Integrating User Affective Signals into Ranking Systems}

\TITLE{Mood-Aware Music Recommendation: Integrating User Affective Signals into Ranking Systems}

\ARTICLEAUTHORS{%

\AUTHOR{Terence Zeng}
\AFF{University of Illinois Urbana-Champaign, Urbana, Illinois 61801, USA, \EMAIL{terence6@illinois.edu}}

\AUTHOR{Abhishek K. Umrawal\footnote{Abhishek K. Umrawal is the corresponding author.}}
\AFF{University of Illinois Urbana-Champaign, Urbana, Illinois 61801, USA, \EMAIL{aumrawal@illinois.edu}}

} 

\ABSTRACT{%
Recommendation systems are essential in modern music streaming platforms due to the vast amount of available content. While collaborative filtering is widely used to suggest items based on the preferences of others with similar patterns, it performs poorly in domains where user-item interactions are sparse, such as music. Content-based filtering is an alternative approach that examines
the qualities of the items themselves. Genre, instrumentation, and lyrics have been explored; however, relatively little attention has been given to emotion recognition. Since a user's emotional state strongly influences their music choice, incorporating mood signals offers a promising direction for personalization. 
In this work, we propose a mood-conditioned ranking framework that integrates user affective signals into the recommendation process via softmax-based sampling in the energy–valence space.
We evaluate the approach via single-blind experiments in which participants compare recommendations from the proposed system against a baseline. The results indicate improved perceived recommendation quality, providing preliminary evidence for the effectiveness of incorporating mood-based inputs into music recommendations.
}






\maketitle


\section{Introduction}
Music recommendation plays a critical role in modern streaming platforms, which host massive catalogs and serve hundreds of millions of users worldwide \cite{IFPI2026}. Traditional recommender systems rely heavily on collaborative filtering (CF) \cite{Goldberg,Koren2009}, which generates recommendations based on user-interaction data by matching users with similar taste patterns. However, CF suffers from significant limitations in the music domain. Interaction data often exceeds 99.9\% sparsity \cite{Dror}, platforms suffer from popularity bias \cite{liu2025unbiased}, and there is a lack of explicit feedback, such as numerical ratings \cite{hu2008collaborative}.

Content-based filtering provides an alternative by leveraging item attributes, such as audio characteristics, rather than relying solely on user-item interactions. In the music domain, early systems such as MARSYAS \cite{Tzanetakis} and feature representations based on Mel-frequency Cepstral Coefficients (MFCCs) \cite{Logan} laid the foundation for extracting both low-level spectral features and higher-level semantic descriptors. However, standard content-based approaches typically assume static user preferences and do not account for the dynamic and context-dependent nature of music consumption. 
In practice, listening behavior is shaped by contextual factors such as activity, time, and user emotion. Prior work shows that streaming environments increase exploration and diversity \cite{datta2018changing}, motivating the use of mood to capture short-term preference shifts beyond long-term taste profiles.

In this work, we propose a mood-aware music recommendation framework that incorporates user-reported affective state into the recommendation process using the energy–valence representation. The core contributions are:
\begin{enumerate}
    \item [(1)] We introduce a method for incorporating user-reported mood into recommendations using the energy–valence representation and a probabilistic selection mechanism.
    \item [(2)] We implement a working system that integrates behavioral data, semantic similarity, and affective features from multiple real-world APIs.
    \item [(3)] We conduct a preliminary user study showing that mood-aware recommendations improve perceived recommendation quality relative to a similarity-based baseline.
\end{enumerate}

More generally, this setting highlights the role of incorporating short-term user context into recommendation systems that are typically based on longer-term preference signals.

\vspace{.1in}
\section{Related Work}
There is a large literature on content-based music recommendation. We focus here on key strands and refer to \cite{zeng2025content} for a detailed review.
\subsection{Audio Signal Analysis and Perceptual Features}
Content-based music recommendation historically relies on audio signal processing. Early research established genre classification methods which used timbre, rhythm, and pitch as features \cite{tzanetakis_genre, knn}. These evolved through time-modulation analysis \cite{panagakis} and deep learning architectures such as Convolutional Neural Networks (CNNs) \cite{liu, duan} and capsule networks \cite{ba} to achieve high classification accuracy. Similarly, instrument detection in polyphonic audio has progressed from Mel-Frequency Cepstral Coefficients (MFCCs) \cite{eronen, rabiner} to multi-label deep classifiers \cite{solanki}. These acoustic qualities are often aggregated into high-level perceptual features such as danceability, valence, and energy \cite{pastukhov}, which correlate strongly with human emotional perception \cite{gerhard, Panda}.
In Music Emotion Recognition (MER), models use either categorical taxonomies \cite{HuXiao, Klimmt} or dimensional spaces, such as Russell's valence-arousal circumplex \cite{russell}. While dimensional models enable continuous mapping of emotional expressions, hybrid approaches combining discrete categories with continuous variables have shown superior effectiveness in personalization \cite{Kang, Roda}.
\subsection{Lyrical Semantic Analysis}
Beyond raw audio signals, textual lyrics provide direct semantic context for emotion recognition. Lyrical analysis is computationally lightweight compared to signal processing \cite{vystrvcilova}, but early bag-of-words and latent semantic indexing methods \cite{logan_semantic} achieved lower accuracy than audio-based approaches. Modern systems leverage hybrid representations combining audio and text embeddings to project songs onto the arousal-valence plane \cite{zhang, schaab}. Furthermore, Large Language Models (LLMs) have enabled the generation of semantic summaries to bypass length and copyright constraints in recommender pipelines \cite{tekle}, as well as the creation of emotion-aware song lyrics \cite{ding}.
\subsection{Context-Aware and Demographic Personalization}
As user preferences are dynamic, context-aware recommender systems use situational variables such as time of day, current activity, and environment \cite{kim_context, kaminskas}. Streaming frameworks have proposed passive context inference via environmental audio capture \cite{Hulaud} or conversational voice interaction \cite{wirfs}. Demographic attributes, including age, gender, and nationality, also predict music exploration profiles and preference patterns \cite{krismayer, vigliensoni}, with computer-vision models even enabling real-time audience estimation in public spaces \cite{mammadli}. 

While music emotion recognition is well-studied for labeling tracks, dynamically matching user-reported emotional states to music filtering remains less explored. This work bridges this gap by combining traditional track similarity with mood-based filtering.

\vspace{.1in}
\section{Methodology}


We consider the problem of recommending songs conditioned on a user’s current mood. We formalize this setting as a probabilistic mood-conditioned ranking problem, where candidate items are scored based on proximity in affective space and sampled via a softmax (Boltzmann) distribution. To personalize recommendations according to user affective state, we represent both user mood and audio content in a shared energy–valence space and select candidate songs based on proximity to a target mood. This enables the incorporation of short-term user context into the recommendation process while operating over a fixed candidate pool derived from historical preferences.

\subsection{Energy-Valence Model}
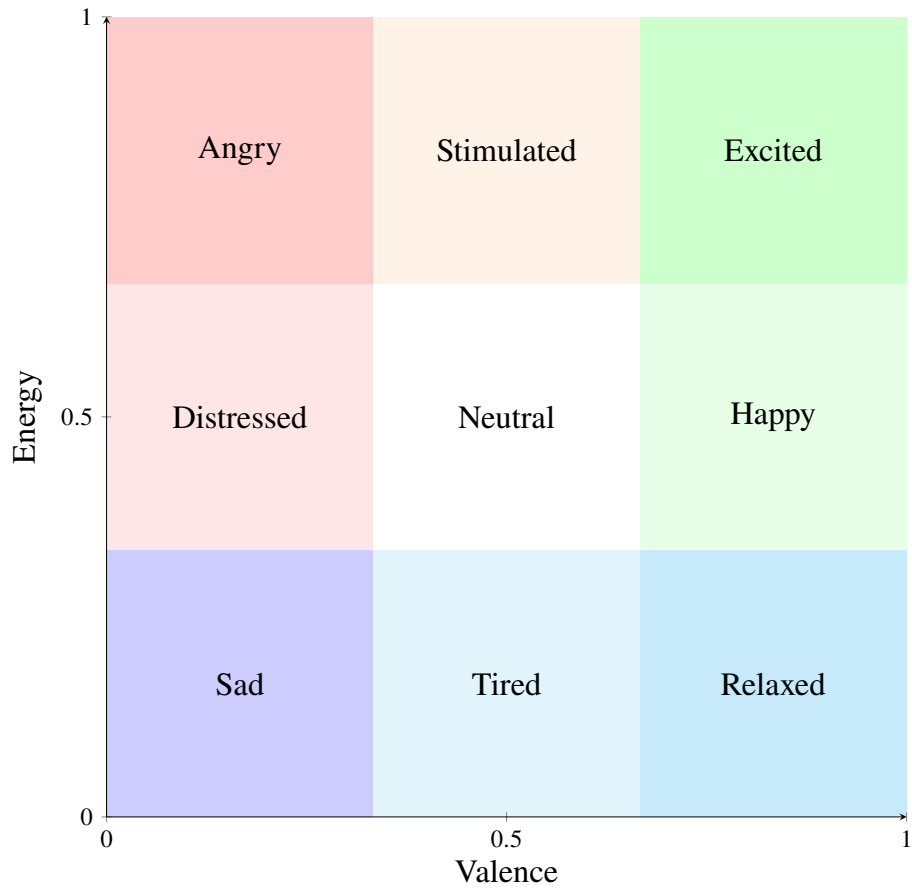
\begin{figure}[h!]
    \centering
    \scalebox{.85}{
    \begin{tikzpicture}
    \begin{axis}
    [axis lines=left, xlabel=\large{Valence}, ylabel=\large{Energy},
    xmin = 0, xmax = 1, ymin = 0, ymax = 1,
    xtick = {0, 0.5, 1}, ytick = {0, 0.5, 1},
    width=0.85\columnwidth, height=0.85\columnwidth,
    axis on top=true]

    \fill[blue!20]      (axis cs:0,0) rectangle (axis cs:1/3,1/3);
    \fill[cyan!10]      (axis cs:1/3,0) rectangle (axis cs:2/3,1/3);
    \fill[cyan!20]      (axis cs:2/3,0) rectangle (axis cs:1,1/3);
    \fill[red!10]       (axis cs:0,1/3) rectangle (axis cs:1/3,2/3);
    \fill[white!10]     (axis cs:1/3,1/3) rectangle (axis cs:2/3,2/3);
    \fill[green!10]     (axis cs:2/3,1/3) rectangle (axis cs:1,2/3);
    \fill[red!20]       (axis cs:0,2/3) rectangle (axis cs:1/3,1);
    \fill[orange!10]    (axis cs:1/3,2/3) rectangle (axis cs:2/3,1);
    \fill[green!20]     (axis cs:2/3,2/3) rectangle (axis cs:1,1);
    
    \node at (axis cs:1/6, 1/6) {\large{Sad}};
    \node at (axis cs:1/6, 1/2) {\large{Distressed}};
    \node at (axis cs:1/6, 5/6) {\large{Angry}};
    \node at (axis cs:1/2, 5/6) {\large{Stimulated}};
    \node at (axis cs:5/6, 5/6) {\large{Excited}};
    \node at (axis cs:5/6, 1/2) {\large{Happy}};
    \node at (axis cs:5/6, 1/6) {\large{Relaxed}};
    \node at (axis cs:1/2, 1/6) {\large{Tired}};
    \node at (axis cs:1/2, 1/2) {\large{Neutral}};
    \end{axis}
    \end{tikzpicture}
    }
    \caption{Mapping of mood categories onto the energy-valence model, adapted from Thayer \cite{thayer1990biopsychology}.}
    \label{fig:mood-model}
\end{figure}

The Energy-Valence Model, proposed by Russell \cite{russell}, defines human emotional state along two orthogonal dimensions in the energy–valence space. Energy (or arousal) measures alertness and stimulation, while valence measures emotional pleasantness. Both attributes are normalized to the range $[0, 1]$. Following Thayer \cite{thayer1990biopsychology}, various mood categories map to distinct regions on this plane (e.g., Excited in high-energy/high-valence, Sad in low-energy/low-valence), as shown in Figure \ref{fig:mood-model}. Audio tracks can also be mapped onto this plane using features extracted via signal processing and learning-based models \cite{yang2008regression}.

\subsection{Song Selection Models}
Given a target user mood $(v, e)$ and a candidate song pool $A[1,\ldots,n]$ with valence and energy arrays $V[1,\ldots,n]$ and $E[1,\ldots,n]$, we consider two selection methods.

\subsubsection{$k$-Nearest Neighbors (Baseline)}
Our baseline approach is adapted from the $k$-nearest neighbor algorithm \cite{peterson2009k}. We select the $k$ closest songs in the candidate pool to the user’s target mood. Distance is defined as the squared Euclidean distance:
\begin{equation}
D[i] = (V[i] - v)^2 + (E[i] - e)^2
\end{equation}
Songs are sorted in ascending order of $D[i]$, and the top $k$ are returned.

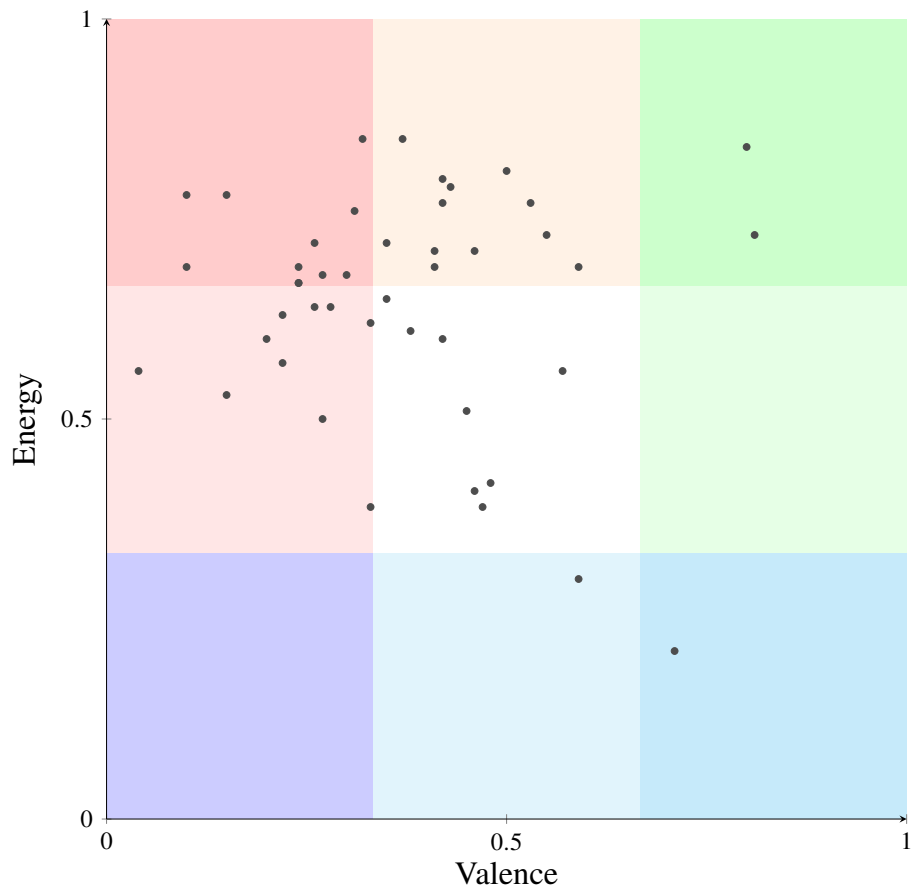
\begin{figure}[h!]
    \centering
    \scalebox{.85}{
    \begin{tikzpicture}
    \begin{axis}
    [axis lines=left, xlabel=\large{Valence}, ylabel=\large{Energy},
    xmin = 0, xmax = 1, ymin = 0, ymax = 1,
    xtick = {0, 0.5, 1}, ytick = {0, 0.5, 1},
    width=0.85\columnwidth, height=0.85\columnwidth,
    axis on top=true,
    only marks,
    mark size=1.5pt]
    \fill[blue!20]      (axis cs:0,0) rectangle (axis cs:1/3,1/3);
    \fill[cyan!10]      (axis cs:1/3,0) rectangle (axis cs:2/3,1/3);
    \fill[cyan!20]      (axis cs:2/3,0) rectangle (axis cs:1,1/3);
    \fill[red!10]       (axis cs:0,1/3) rectangle (axis cs:1/3,2/3);
    \fill[white!10]     (axis cs:1/3,1/3) rectangle (axis cs:2/3,2/3);
    \fill[green!10]     (axis cs:2/3,1/3) rectangle (axis cs:1,2/3);
    \fill[red!20]       (axis cs:0,2/3) rectangle (axis cs:1/3,1);
    \fill[orange!10]    (axis cs:1/3,2/3) rectangle (axis cs:2/3,1);
    \fill[green!20]     (axis cs:2/3,2/3) rectangle (axis cs:1,1);
    \addplot [
        mark=*,
        color=black!70
    ] coordinates {
        (0.33, 0.62)
        (0.24, 0.67)
        (0.24, 0.67)
        (0.37, 0.85)
        (0.22, 0.57)
        (0.28, 0.64)
        (0.81, 0.73)
        (0.45, 0.51)
        (0.41, 0.71)
        (0.59, 0.30)
        (0.42, 0.77)
        (0.38, 0.61)
        (0.31, 0.76)
        (0.10, 0.78)
        (0.71, 0.21)
        (0.35, 0.72)
        (0.41, 0.69)
        (0.53, 0.77)
        (0.10, 0.69)
        (0.42, 0.60)
        (0.26, 0.72)
        (0.27, 0.50)
        (0.42, 0.80)
        (0.20, 0.60)
        (0.15, 0.78)
        (0.46, 0.41)
        (0.43, 0.79)
        (0.30, 0.68)
        (0.33, 0.39)
        (0.80, 0.84)
        (0.35, 0.65)
        (0.57, 0.56)
        (0.55, 0.73)
        (0.22, 0.63)
        (0.24, 0.69)
        (0.50, 0.81)
        (0.27, 0.68)
        (0.04, 0.56)
        (0.32, 0.85)
        (0.47, 0.39)
        (0.26, 0.64)
        (0.46, 0.71)
        (0.48, 0.42)
        (0.15, 0.53)
        (0.59, 0.69)
    };
  
    \end{axis}
    \end{tikzpicture}
    }
    \caption{Distribution of an example user's top listening history tracks on the energy-valence plane.}
    \label{fig:song-graph-general}
\end{figure}

As shown in Figure \ref{fig:song-graph-general}, an example user's listening history is distributed across the valence–energy plane. Under a deterministic selection scheme such as KNN, repeated queries for a fixed emotional target will return the same set of songs. This provides limited utility, especially since a user's short-term mood fluctuates while their overall music collection remains fixed. This limitation motivates a probabilistic approach to introduce recommendation diversity.

\subsubsection{Softmax/Boltzmann Selection}

To balance alignment with the target mood and diversity in recommendations, we propose a Softmax selection method (Algorithm \ref{alg:softmax}) based on the Boltzmann distribution \cite{softmax}. We compute the squared Euclidean distance $D[i]$ for all candidates and map them to unnormalized Boltzmann weights:
\begin{equation}
W[i] = \exp\left( - \frac{D[i]}{k} \right)
\end{equation}
where the decay factor $k$ controls selection sharpness. We then normalize the weights to form a probability distribution:
\begin{equation}
P[i] = \frac{W[i]}{\sum_{j=1}^n W[j]}
\end{equation}
Finally, we sample $r$ distinct tracks according to $P$. This ensures that songs closer to the target mood are favored, while less similar but novel recommendations retain a non-zero probability of being selected.

\begin{algorithm}[!htbp]
    \caption{Softmax Song Selection}
    \label{alg:softmax}
    \begin{algorithmic}
        \Require user mood $(v, e)$, decay factor $k$, number of recommendations $r$, songs $A[1,\ldots,n]$ with valences $V[1,\ldots,n]$ and energies $E[1,\ldots,n]$
        \Statex
        \State $D[1,\ldots,n]$ \Comment{distance between each song and user mood}
        \ForAll{$i \in [1,n]$}
            \State $D[i] \gets (V[i] - v)^2 + (E[i] - e)^2$
        \EndFor
        \Statex
        \State $W[1,\ldots,n]$ \Comment{unnormalized Boltzmann weights}
        \ForAll{$i \in [1,n]$}
            \State $W[i] \gets \exp\!\left(-\frac{D[i]}{k}\right)$
        \EndFor
        \Statex
        \State $P[1,\ldots,n]$ \Comment{normalized probabilities}
        \State $Z \gets \sum_{i=1}^{n} W[i]$
        \ForAll{$i \in [1,n]$}
            \State $P[i] \gets W[i] / Z$
        \EndFor
        \Statex
        \State Sample $r$ distinct indices according to probability distribution $P$ 
        \State \Return corresponding songs from $A$
    \end{algorithmic}
\end{algorithm}

\vspace{.1in}
\section{Experiments}

To evaluate our mood-assisted recommendation framework against a traditional baseline under realistic conditions, we developed a real-time web application to facilitate data collection and user trials.

\subsection{System Architecture and Data Pipeline}
\begin{figure}[h!]
    \centering
    \includegraphics[width=.93\columnwidth]{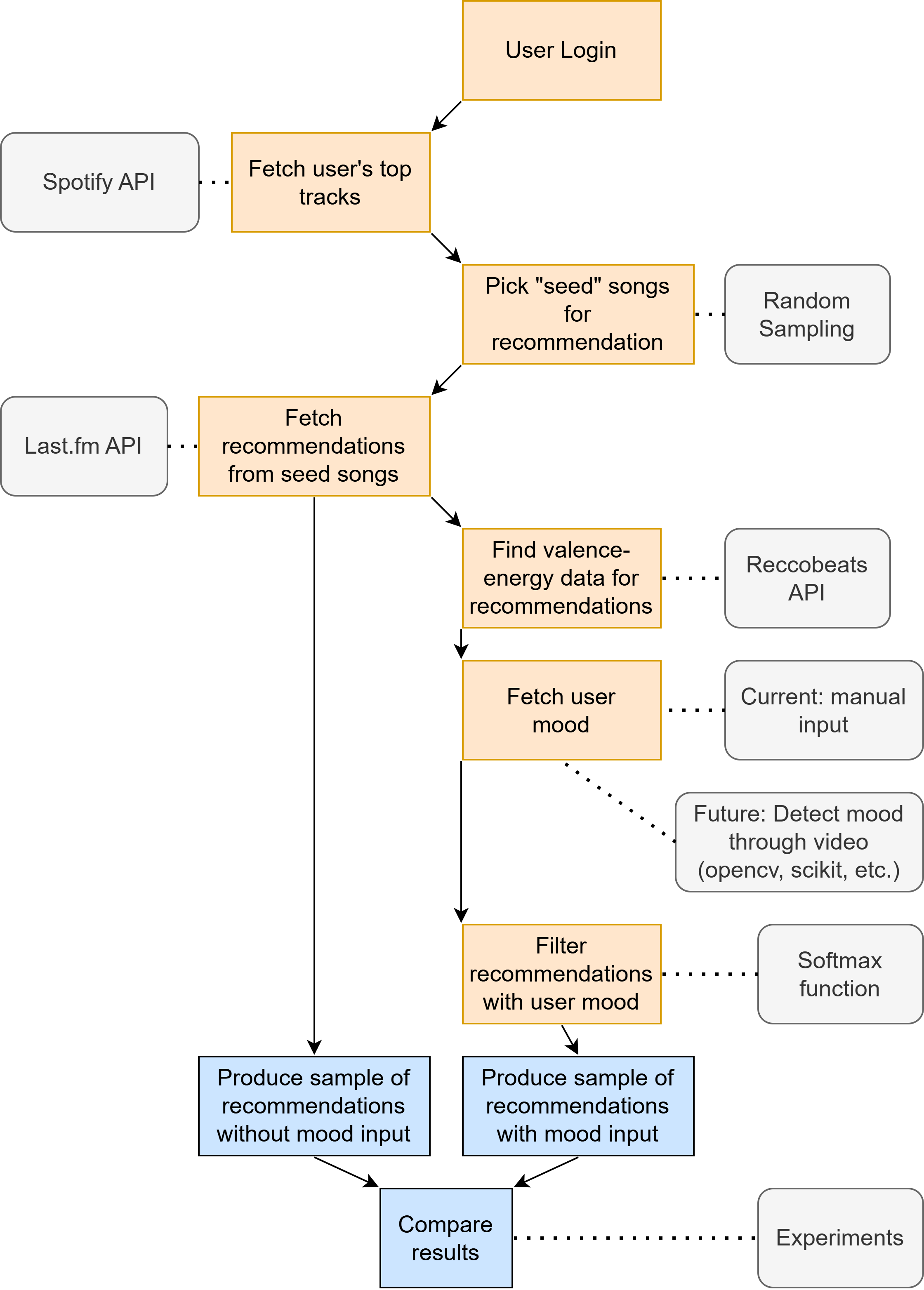}
    \caption{Flowchart outlining the experimental data processing and recommendation pipeline.}
    \label{fig:methods_flowchart}
\end{figure}

The recommendation pipeline consists of three sequential stages, as illustrated in Figure \ref{fig:methods_flowchart}:

\begin{enumerate}
    \item \textbf{User Profiling (Spotify API):} Users authenticate via Spotify OAuth. The system queries the Spotify Web API to retrieve the user's top tracks across short, medium, and long-term listening histories. A random subset of these tracks is selected as ``seed songs'' representing the user's base preferences.

    \item \textbf{Candidate Exploration (Last.fm API):} For each seed track, we query the Last.fm \texttt{track.getSimilar} API to retrieve similar tracks. This step generates a diverse candidate pool of tracks that are semantically aligned with the user's historical preferences, mitigating cold-start limitations in collaborative approaches. Since Last.fm identifiers are not directly compatible, candidate track metadata (names and artists) are matched to Spotify IDs using the Spotify search endpoint.

    \item \textbf{Audio Feature Retrieval (ReccoBeats API):} For each candidate track, the system retrieves valence and energy values from the ReccoBeats API.
\end{enumerate}

Once the candidate pool is populated with features, the system generates two playlists: (1) a \text{control} recommendation based on similarity metrics ($k$-nearest neighbors in the energy–valence space); and (2) a \text{mood-assisted} recommendation filtered toward a user-selected target mood using Softmax Song Selection (Algorithm \ref{alg:softmax}). To reduce latency, similarity queries and feature retrieval are executed concurrently using Python's asynchronous I/O framework.

\subsection{Experimental Design}
We conducted a single-blind user study. Participants logged into the web platform, which automatically retrieved their top tracks and generated candidate pools. Participants then selected their current desired mood from nine options corresponding to Thayer’s mood quadrants. The system presented two anonymous recommendations: one from the baseline control model and one from the mood-assisted model. Participants rated each recommendation on a 5-point Likert scale (1 = Poor, 5 = Excellent) and optionally provided comments. A total of 27 evaluation pairs from 6 participants were collected.

\section{Results}

We evaluate the recommendation quality by analyzing user ratings across the control and mood-assisted models.

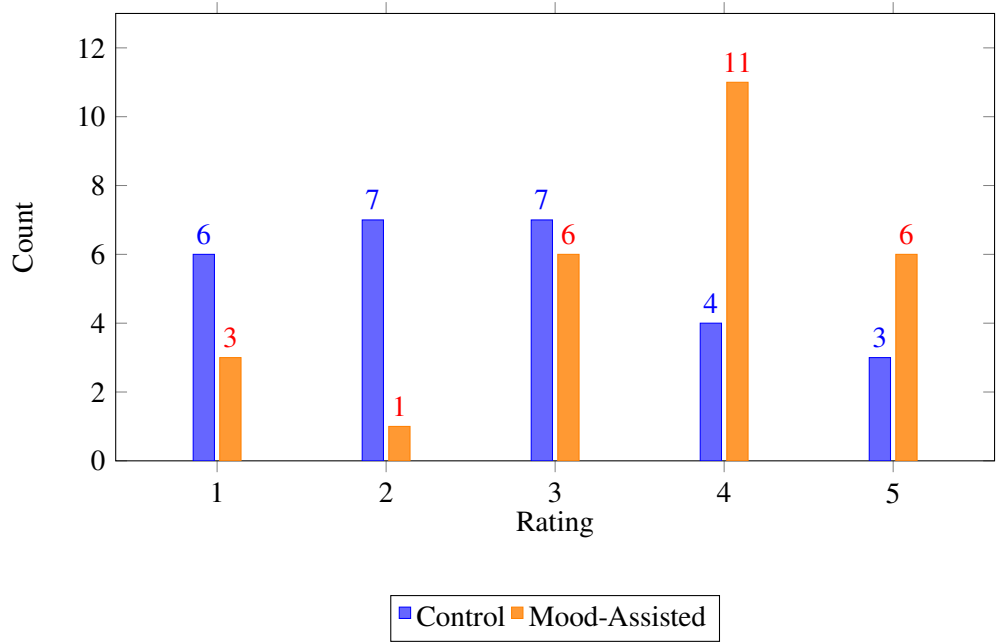
\begin{figure}[h]
\centering
\begin{tikzpicture}
\begin{axis}[
    ybar,
    bar width=8pt,
    width=.8\columnwidth,
    height=7.5cm,
    ymin=0,
    ymax=13,
    enlarge x limits=0.15,
    legend style={at={(0.5,-0.3)},anchor=north,legend columns=2},
    symbolic x coords={1,2,3,4,5},
    xtick=data,
    xlabel={Rating},
    ylabel={Count},
    nodes near coords,
    nodes near coords align={vertical},
    legend image code/.code={
        \draw[#1,fill=#1] (0cm,0cm) rectangle (0.15cm,0.15cm);
    }
]

\addplot+[fill=blue!60, draw=blue] coordinates {
    (1,6)
    (2,7)
    (3,7)
    (4,4)
    (5,3)
};

\addplot+[fill=orange!80, draw=orange] coordinates {
    (1,3)
    (2,1)
    (3,6)
    (4,11)
    (5,6)
};

\legend{Control, Mood-Assisted}

\end{axis}
\end{tikzpicture}
\caption{Distribution of participant ratings on a 5-point scale ($n=27$).}
\label{fig:rating-graph}
\end{figure}

\subsection{Statistical Analysis}

Across all trials, the mood-assisted recommendation model consistently outperformed the control model. Mood-assisted recommendations received a mean rating of $3.59$ (SD = 1.19), whereas control recommendations received a mean rating of $2.67$ (SD = 1.28). As illustrated in Figure \ref{fig:rating-graph}, the score distribution for the mood-assisted system shows a clear upward shift, with 63\% of recommendations rated as 4 or 5, compared to only 26\% for the control system.

To evaluate statistical significance, we performed a Mann–Whitney $U$ test \cite{MannWhitney}, appropriate for ordinal, non-normal Likert ratings. With sample sizes $n_1 = n_2 = 27$, the test yields a $z$-score of $-2.56$, corresponding to a two-tailed $p$-value of approximately 0.01. This indicates a statistically significant improvement in user-perceived recommendation quality when incorporating mood-based inputs.

\subsection{Discussion}
The results suggest that mapping user-reported moods onto the valence–energy plane provides a meaningful personalization signal beyond track similarity. By combining Last.fm similarity data (capturing preference continuity) with ReccoBeats audio feature filtering (capturing emotional relevance), the system avoids recommending tracks that, while similar in genre or style, conflict with the user's immediate emotional state. 

These findings indicate that user preferences in music recommendation are not fully captured by static historical behavior alone, but also depend on transient contextual factors such as mood. Incorporating such signals can therefore meaningfully alter recommendation outcomes, even when the underlying candidate pool remains unchanged.

However, the efficacy of the mood-assisted system varies across specific mood targets (Table \ref{table:mood-results}). The most substantial improvements occur for ``Relaxed'' (+2.67) and ``Sad'' (+2.00) moods. Conversely, for ``Stimulated'' and ``Distressed'' moods, no improvement is observed. This suggests that users seeking high-arousal or high-tension music may have narrower or more variable interpretations of emotional categories, or that baseline recommendations already align sufficiently with these states.
\begin{table}[h!]
\centering
\caption{Mean ratings grouped by target mood.}
\label{table:mood-results}
\begin{tabular}{lcc}
\hline
\textbf{Target Mood} & \textbf{Control} & \textbf{Mood-Assisted} \\
\hline
Relaxed    & 2.33 & 5.00 \\
Sad        & 2.00 & 4.00 \\
Tired      & 2.33 & 3.67 \\
Distressed & 2.67 & 2.67 \\
Neutral    & 3.33 & 3.67 \\
Happy      & 3.67 & 4.00 \\
Angry      & 1.00 & 2.00 \\
Stimulated & 4.67 & 4.67 \\
Excited    & 2.00 & 2.67 \\
\hline
\end{tabular}
\end{table}
This study provides preliminary evidence that incorporating mood-aware filtering improves recommendation quality; future work will validate these findings on a larger scale. 
While promising, several limitations remain:
(1) The user study relies on a small participant cohort (27 evaluation pairs), which limits statistical power and generalizability. 
(2) Manual self-reporting introduces subjective bias, as users’ perception of their mood and its mapping to discrete labels may vary. 
(3) Spotify-to-Last.fm track reconciliation is imperfect, occasionally excluding less popular tracks due to missing cross-platform metadata, and thereby constraining the candidate pool.

\vspace{.1in}

\section{Conclusion and Future Work}

This work demonstrates that incorporating user mood via the energy–valence plane is an effective approach for enhancing personalized music recommendations. By combining similarity-based candidate retrieval with softmax-weighted emotion filtering, our framework aligns recommendations with users’ short-term emotional states while preserving novelty. A user study provides preliminary evidence of improved perceived recommendation quality over baseline methods. These findings highlight the importance of incorporating short-term contextual signals into recommendation systems traditionally driven by long-term preferences.

Future work will focus on addressing the limitations of the current study and extending the modeling framework. 
First, we plan to validate the approach at larger scale through broader user studies and longitudinal evaluation. 
Second, we aim to extend emotional modeling beyond the two-dimensional energy–valence plane to capture more nuanced affective and cognitive states. 
Third, we will explore multimodal user signals (e.g., passive sensor data, facial expressions, or contextual cues) to enable automatic mood inference and reduce reliance on self-reported inputs. 
Finally, improving cross-platform track matching and expanding candidate generation pipelines will further enhance coverage and recommendation diversity.



\vspace{.1in}
\bibliographystyle{informs2014} 
\bibliography{bibliography} 






  



\end{document}